# Phase-matching of high harmonic generation in twisted solids

Chenjun Ma[1#], Chen Huang[1#], Yilong You[1#], Huazhan Liu[1], Zhitong Ding[1], Mingchao Ding[2], Jin Zhang[3], Guixin Li[4], Zhipei Sun[5], Shiwei Wu[6], Chaojie Ma[1]*, Enge Wang[7,8]*, Hao Hong[1,9]* and Kaihui Liu[1,7]*

[1] State Key Lab for Mesoscopic Physics, Frontiers Science Centre for Nano-optoelectronics, School of Physics, Peking University, Beijing, China

[2] Institute of Physics, Chinese Academy of Sciences, Beijing, China

[3] National Center for Nanoscience and Technology, Chinese Academy of Sciences, Beijing, China

[4] Department of Materials Science and Engineering, Southern University of Science and Technology, Shenzhen, China

[5] QTF Centre of Excellence, Department of Electronics and Nanoengineering, Aalto University, Espoo, Finland

[6] State Key Laboratory of Surface Physics and Department of Physics, Fudan University, Shanghai, China

[7] International Centre for Quantum Materials, Collaborative Innovation Centre of Quantum Matter, Peking University, Beijing, China

[8] Tsientang Institute for Advanced Study, Zhejiang, China

[9] Interdisciplinary Institute of Light-Element Quantum Materials and Research Centre for Light-Element Advanced Materials, Peking University, Beijing, China

[#] These authors contributed equally to this work

\* Correspondence: khliu@pku.edu.cn, haohong@pku.edu.cn, chaojiema@pku.edu.cn, egwang@pku.edu.cn

**High harmonic generation (HHG) in solids could enable attosecond and ultraviolet light sources with high compactness, great controllability and rich functions. However, the HHG process is accompanied by a quite large wavevector mismatch that is uncompensated by any traditional phase-matching method, directly limiting its energy conversion efficiency. Here, we propose an effective strategy for phase-matching of HHG with arbitrary harmonic orders in solids. Two flakes of solids with an interlayer twist induce a nonlinear optical phase that depends on the crystal symmetry, twist angle and harmonic order, which can be accurately designed to compensate for the phase mismatch in HHG. Guided by the twist-phase-matching theory, we achieved a record-high conversion efficiency of ~$1.5 \times 10^{-5}$ for the fifth HHG in twisted hexagonal boron nitride crystals with a total thickness of only 1 μm. Our work establishes a foundation for developing ultrashort-wavelength and ultrafast-pulse laser sources in compact solid-state tabletop systems for fundamental and applied sciences.**

High harmonic generation (HHG) is a nonlinear optical process in which photons with frequencies that are integer multiples (typically ≥5) of the fundamental wave are emitted when an intense light pulse interacts with matter[1,2]. The electric field of the light is sufficiently strong to rival the atomic binding field of electrons, thus driving electrons in the matter to undergo rapid acceleration and inducing coherent radiation of high-energy photons with well-defined timings, phases and wavelengths related to the fundamental wave. Initially, HHG was observed in plasma generated from aluminium targets[3]. Shortly thereafter, noble gases were identified as useful HHG sources that can operate in the nonperturbative region with high nonlinearity[4-6]. The highly nonlinear nature of HHG compresses electromagnetic pulses to the attosecond timescale and extends the photon energy into the soft X-ray regime, opening the door to attosecond science[7-10] and strong-field optical physics[11-14].

Recent advancements in short-pulse, high-intensity, long-wavelength laser technologies—which are capable of generating strong electric fields while minimizing material damage—have enabled a shift in the research focus towards solid-state systems, such as wide bandgap semiconductors[15-18], topological materials[19,20] and two-dimensional (2D) materials[21-24]. Compared with gas-state systems, solids exhibit unique characteristics, including densely packed atomic structures, lattice periodicity and geometric symmetry. The high atomic density of solids (typically 3 to 6 orders of magnitude greater than that of gases) facilitates excitation of more electrons per unit volume and significantly enhances harmonic generation processes. The lattice periodicity in solids introduces continuous electronic band structures, enabling an additional HHG pathway through intraband currents alongside interband polarization[25-27]. Consequently, solid-state systems have the potential to produce ultrastrong HHG outputs. However, the energy conversion efficiency of HHG in solids remains considerably low ($\sim 10^{-8}$) compared to that in gas ($\sim 10^{-5}$), primarily due to the lack of an effective phase-matching strategy to coherently enhance the HHG.

Phase-matching involves establishing a proper phase relationship between the fundamental and generated waves to maximize the optical parametric process efficiency[1, 2]. In gas-state systems, HHG phase-matching is achieved by engineering the gas pressure, species composition, spatial modes, and waveguide diameters to balance the wavevector mismatch arising from neutral dispersion, plasma dispersion, the Gouy phase, and the dipole phase[28-32]. In contrast, for free-space solid-state systems, phase-matching methods have been successfully demonstrated only for low-order (≤3) nonlinear processes. For HHG, birefringent-phase-matching is impractical due to the inability of birefringent coefficients to compensate for the quite large wavevector mismatch. Theoretically, quasi-phase-matching could be applied to the generation of even-order high harmonics; however, the experimental results remain debatable, as direct HHG processes are often indistinguishable from cascaded processes[33-35]. A universal theoretical strategy and experimental demonstration of phase-matching for HHG in solids have yet to be realized.

The inherent geometric symmetry of solids offers novel opportunities for addressing phase mismatch. An interlayer rotation in a 2D crystal has been reported to induce a phase shift in the second-order polarization, which can be applied for second harmonic generation (SHG) phase-matching[24, 36-39]. However, the situation becomes quite complex for HHG, as the HHG susceptibility involves multiple components that complicate the relationship between the twist angle and polarization. Here, we demonstrate that the twist-induced phase can be simplified to a Pancharatnam-Berry (P-B) phase that can be designed for phase-matching in HHG with arbitrary harmonic orders. Guided by the developed twist-phase-matching theory, we designed a solid HHG source using an assembly of hexagonal boron nitride (hBN) flakes and achieved a record-high fifth HHG efficiency of ~$1.5\times10^{-5}$ in a 1 μm thick crystal.

We began by considering two ultrathin flakes assembled with an interfacial twist angle of $\theta$ for HHG (Fig. 1a). The polarization states of the $n$th-order HHG signals from these two flakes are labelled as $|a\rangle$ and $|b\rangle$, corresponding to points A and B on the Poincaré sphere, respectively (Fig. 1b). $|a\rangle$ and $|b\rangle$ are linked by the twist operation **J**, and have a phase difference $\phi$ that can be depicted as[40-42]

$$\phi = \arg\langle a|b\rangle = \arg\langle a|\mathbf{J}|a\rangle = \arg(\det \mathbf{J})/2 + [\Omega_{ABQ} - \Omega_{BA(-Q)}]/4, \quad (1)$$

where Q is a point on the Poincaré sphere that related to matrix **J** and $\Omega_{ABQ}$ and $\Omega_{BA(-Q)}$ are solid angles of the spherical triangles formed by ABQ and BA(−Q), respectively. In Eq. (1), the first term is a dynamic phase, and the second term is known as the P-B phase $\phi_{PB}$.

Typically, the optical nonlinear susceptibility contains multiple tensor components that differ in crystals with different symmetries. For simplification, we consider only the pure circularly polarized components, and the HHG polarization can be expressed as $P(n\omega) = \varepsilon_0 \chi_1^{(n)} E_\sigma^n$ ($\sigma = \pm 1$ represents the circular polarization state). Thus, Q is fixed as [0, 0, 1], the dynamic phase becomes zero, and the P-B phase can be derived as

$$\tan\phi_{PB} = \tan(n \pm 1)\theta \cdot \overrightarrow{OQ} \cdot \overrightarrow{OA}, \quad (2)$$

where O is the centre point of the Poincaré sphere. Furthermore, if $A$ is circular polarization state and $|\overrightarrow{OQ} \cdot \overrightarrow{OA}| = 1$, Eq. 2 evolves into its most simplified form of $\phi_{PB} = (n \pm 1)\theta$. At this point, the original spherical quadrilateral degenerates into the region enclosed by two meridians on the Poincaré sphere (Fig. 1b).

P-B phase $\phi_{PB}$ is determined by not only the twist angle $\theta$, but also the rotational symmetries of the used nonlinear optical material[38]. The conservation of spin angular momentum in nonlinear interaction states a selection rule that an $m$-fold rotationally symmetric material only allows harmonic orders of $n = lm \mp 1$ ($l$ is an arbitrary integer, and the "−" and "+" signs correspond to HHG light with the opposite and same circular polarization state as the fundamental wave, respectively). Therefore, $\phi_{PB} = lm\theta$ depends on the crystal rotational symmetry, twist angle and harmonic order.

Experimentally, a crystalline 2D hBN multilayer was selected as the model platform for achieving the twist-phase-matching design of HHG because of its large optical nonlinearity, broad bandgap, excellent physicochemical stability and high laser damage threshold. The multilayers exhibit an antiparallel-stacking interlayer configuration and belong to the $D_{6h}$ space group (Fig. 2a). Under excitation at ~1750 nm, the HHG spectrum of a single hBN flake exhibits three peaks at 583 nm, 350 nm, and 250 nm, corresponding to odd-order harmonic waves of the third, fifth, and seventh harmonic generations, respectively (Fig. 2b). Notably, the intensity of the fifth HHG is significantly high, approaching ~1/30 of that of the third harmonic generation. Following careful calibration, the nonlinear susceptibility of hBN is estimated to be $\chi^{(5)} \approx 1 \times 10^{-39}$ m$^4$/V$^4$ (two orders of magnitude greater than that of silica, Supplementary Fig. 1). For a crystal with the $D_{6h}$ space group, there are two independent tensor components ($\chi_1^{(5)}$ and $\chi_2^{(5)}$) in the fifth-order nonlinear susceptibility, and the fifth harmonic nonlinear polarization can be written as

$$\begin{cases} P_-(5\omega) = \varepsilon_0\left(\chi_1^{(5)}E_+^5 + 10\chi_2^{(5)}E_+^2 E_-^3\right) \\ P_+(5\omega) = \varepsilon_0\left(\chi_1^{(5)}E_-^5 + 10\chi_2^{(5)}E_+^3 E_-^2\right), \end{cases} \quad (3)$$

where $\chi_1^{(5)}$ responses for the pure circularly polarized light excitation, and $\chi_2^{(5)}$ responses for the elliptically polarized light excitation. For linearly polarized excitation with $E_- = \frac{1}{\sqrt{2}}E_0 e^{-i\alpha}$ and $E_+ = \frac{1}{\sqrt{2}}E_0 e^{i\alpha}$, the fifth HHG output intensity can be written as

$$\begin{aligned} I(5\omega) \propto |P(5\omega)|^2 &= |P_-(5\omega)|^2 + |P_+(5\omega)|^2 \\ &= \frac{E_0^{10}}{16}\varepsilon_0^2\left((\chi_1^{(5)})^2 + 100(\chi_2^{(5)})^2 + 20\chi_1^{(5)}\chi_2^{(5)}\cos 6\alpha\right), \end{aligned} \quad (4)$$

where $\alpha$ is the intersection angle between the fundamental wave polarization and the armchair direction of the hBN crystal (Fig. 2c). This equation very well describes the six-petal pattern observed in our polarization-dependent HHG experiment (Fig. 2d), which revealed that the ratio of these two tensor components is $\chi_1^{(5)}/\chi_2^{(5)} = 0.7$.

In the following, we applied the P-B phase introduced at the twisted interface to design phase-matching in two hBN flakes. Given the sixfold rotational symmetry of the hBN multilayer, an interfacial rotation by an angle $\theta$ imparts a $6\theta$ P-B phase to the fifth HHG (Fig. 3a). Consequently, the nonlinear polarization evolves as

$$\begin{cases} P'_-(5\omega) = \varepsilon_0(\chi_1^{(5)}E_+^5 e^{-i6\theta} + 10\chi_2^{(5)}E_+^2 E_-^3) \\ P'_+(5\omega) = \varepsilon_0(\chi_1^{(5)}E_-^5 e^{i6\theta} + 10\chi_2^{(5)}E_+^3 E_-^2). \end{cases} \quad (5)$$

As the rotation operation and the twisted interface only introduce a phase shift in the fifth harmonic polarization without affecting the fundamental wave, a phase-matching condition can be achieved when $5k_\omega + 6\theta/t = k_{5\omega}$ (Fig. 3b), where $t$ is the thickness of a single hBN flake and $k_\omega$ and $k_{5\omega}$ are the wavevectors of the fundamental and fifth harmonic waves, respectively. Thus, $\theta$ scales linearly with $t$ in our twist-phase-matching theory (Fig. 3c). For a stacking thickness $t$ of the coherence length $l_c = \frac{\lambda(\omega)}{10|n(\omega) - n(5\omega)|}$ (where $n(\omega)$ and $n(5\omega)$ are the refractive indices of the fundamental and fifth harmonic waves), $\theta$ approaches the largest absolute value of the twist angle in a $C_6$ symmetry crystal of $30°$.

The physical picture of twist-phase-matching can be understood from the complex amplitude diagram depicted in Fig. 3d. The fifth harmonic electric field generated by each hBN flake contributes to an arc chord vector, whose magnitude is determined by the flake thickness. The sum of the vectors represents the overall HHG response of the twisted stacked crystal. Twist-phase-matching is a strategy that aligns these vectors in the same direction by introducing the P-B phase at the interface, thereby maximizing the HHG response. Experimentally, we assembled two hBN flakes, each with a thickness of $l_c \approx 320$ nm, at a twist angle of $0°$ or $30°$, respectively (Fig. 3e). As anticipated, the phase-matched crystal with a twist angle of $30°$ exhibits a fourfold enhancement in HHG compared with the monolayer, whereas the phase-mismatched crystal with a twist angle of $0°$ demonstrates a suppressed response.

The twist-phase-matching design promises a continuous enhancement of the parametric output with the number of stacked flakes (Fig. 4a). In the assembly, each flake had a thickness of $l_c/2 = 160$ nm, and the twist angle between adjacent flakes was $-15°$, which satisfies the twist-phase-matching condition. Initially, we tested the excitation-ellipticity-dependent behaviour. As indicated by Eq. (3) for the 1-flake hBN, the output HHG intensity varies with the incident light ellipticity and reaches its maximum under linear polarization with ellipticity = 0 (Fig. 4b, orange curve). As the stacking number increases, the term $\chi_1^{(5)} E_+^5 e^{-i6\theta}$ in Eq. (5) experiences phase-matching, resulting in constructive HHG enhancement; whereas the other polarized components experience phase-mismatching with destructive HHG. Consequently, the ellipticity-dependent fifth HHG curves evolve correspondingly and present a peak at ellipticity = 1 with increasing stacking number, which can be reproduced very well by our simulation. The flake-number-dependent fifth HHG intensity under the twist-phase-matching condition of ellipticity = 1 is shown in Fig. 4c. A quadratic dependence on the flake number is observed, indicating a coherent enhancement of the fifth HHG response between flakes.

The twist-phase-matching in HHG also exhibits an obvious wavelength-dependent behaviour. (Fig. 4d). The six-flake twisted stacked hBN crystal ($t$ = 160 nm, $\theta = -15°$) works at the fundamental excitation wavelength of 1340 nm, as designed. The crystal exhibits a large bandwidth of ~200 nm that is described well by our theoretical model. Finally, we estimated the output HHG efficiency of our twist-phase-matching designed hBN crystal, with a total thickness of 960 nm. The energy conversion efficiency scales quadratically with the excitation power intensity (Fig. 4e). It approaches ~0.0015% before laser-induced crystal damage, representing the highest value among the solid HHG sources ever reported.

In summary, we developed a general phase-matching theory for HHG in twisted solids and achieved a record-high fifth HHG efficiency of ~1.5×10$^{-5}$ in hBN crystals with a thickness of only 1 μm. With the advantage of an ultrathin thickness down to a few micrometres, this material is highly promising for minimizing pulse broadening down to the attosecond time scale. Our study provides a new platform for efficiently generating and engineering HHG light in solids, which should facilitate advances in all-solid compact lasers, on-chip integrated devices and quantum photonic technologies towards the extreme optics regime.

**Acknowledgements:** This work is supported by the National Key R&D Program of China (2022YFA1403500 and 2021YFA1400201), National Natural Science Foundation of China (52025023, 12374167 and 12422406), and the New Cornerstone Science Foundation through the XPLORER PRIZE.

**Competing interests:** The authors declare no competing interests.

**Methods:**

**Sample preparation for twisted hBN flakes.** The hBN flakes were mechanically exfoliated from a commercial bulk hBN crystal. The large flakes (lateral size >100 μm) with specific thicknesses were selected to fabricate the twisted hBN crystal.. The twisted hBN flakes were fabricated using a precisely controlled dry-transfer technique, with a polypropylene carbonate (PPC) film placed on a polydimethylsiloxane (PDMS) stamp. One hBN flake was picked up and transferred onto another, with the rotation angle carefully controlled using a commercial two-dimensional-material rotary transfer stage (JOOIN Technology, 2DT-7-300). The hBN/PPC film was then detached from the stamp by heating to 130 °C. To remove any PPC residue and ensure a better interlayer fit, the twisted hBN flakes were annealed at 500 °C for 8 hours under low-pressure condition (<200 Pa), with Ar (500 sccm) and $H_2$ (50 sccm) gas flow.

**Sample characterization.** Optical images of hBN flakes were taken with an Olympus BX51M microscope. Atomic force microscopic (AFM) mappings were obtained by an Oxford Asylum Research Cypher S system at ambient atmosphere (Supplementary Fig. 3).

Figures and captions:

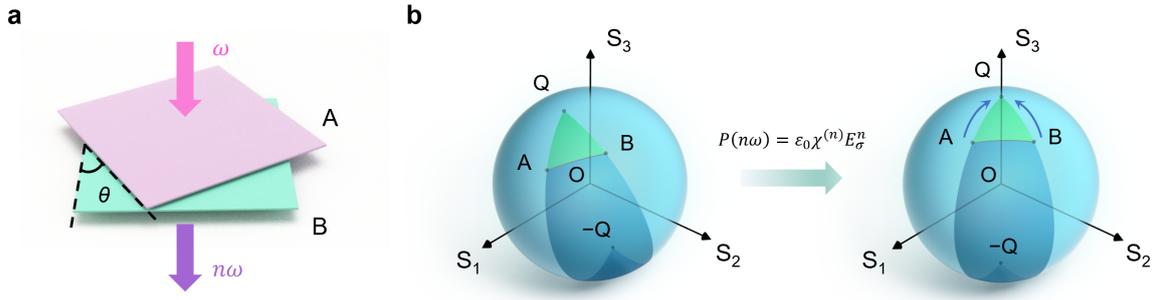

**Fig. 1 | Twist induced Pancharatnam-Berry phase at the interface of two ultrathin solids.**

**a**, Schematic of *n*th-order HHG from two ultrathin solids assembled with an interfacial twist angle of $\theta$. **b**, Illustration of the induced P-B phase on the Poincaré sphere. Points A and B on the Poincaré sphere demonstrate the polarization of HHG light from flakes A and B, respectively. The point Q is related to the twist operation, and −Q is the centrosymmetric point of Q. The phase difference between A and B can be described by the solid angle difference between ABQ and AB(−Q). If circularly polarized laser excitation is considered, points Q, A and B approach the poles, and the P-B phase can be simplified to be $\phi_{PB} = (n \pm 1)\theta$.

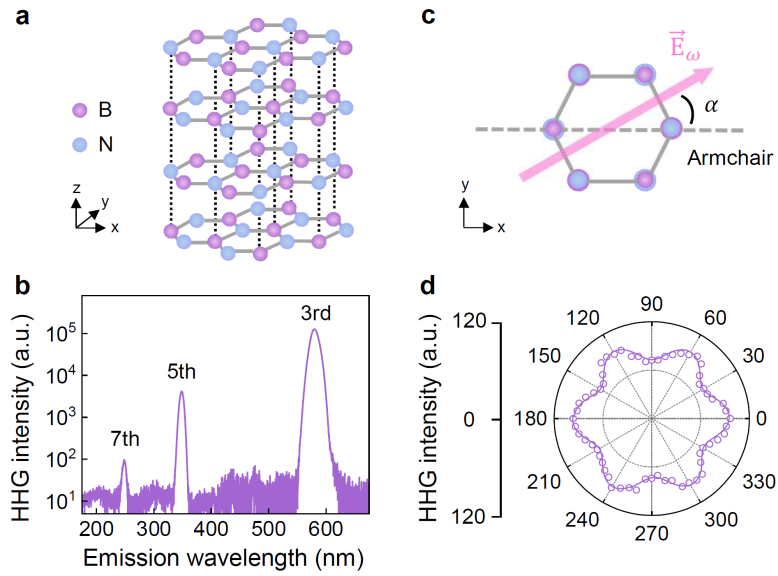

**Fig. 2 | HHG responses of an hBN flake. a**, Schematic of the stacking configuration of hBN. **b**, HHG spectrum of the hBN flake. Under linearly polarized excitation light with a wavelength of 1750 nm and a peak intensity of ~3 TW/cm$^2$, the hBN flake exhibits third, fifth and seventh harmonic responses. **c**, Illustration of the laser polarization orientation relative to the hBN crystalline lattice direction. Here, the laser polarization is rotated $\alpha$ with respect to the fixed the armchair direction of the hBN crystal. **d**, Polarization-dependent HHG spectrum. A six-petal pattern is observed from our measured data (dots) and fitting result (solid curve).

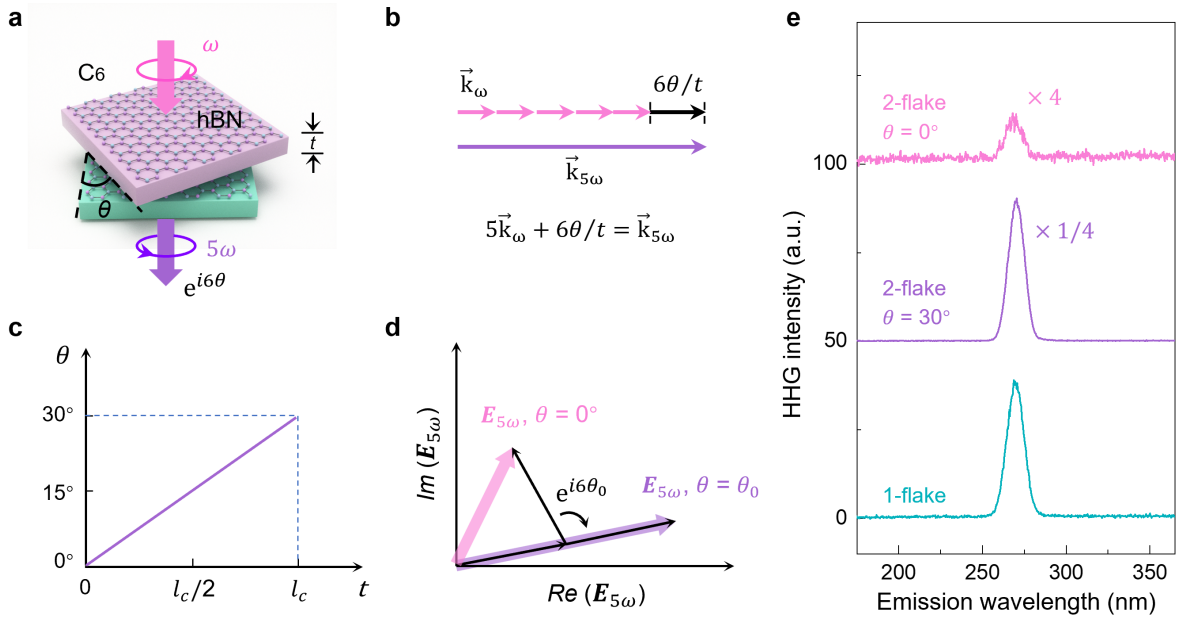

**Fig. 3 | HHG responses of two twisted hBN flakes. a**, Schematic of twisted hBN flakes. A twist angle of $\theta$ introduces a $6\theta$ P-B phase into the fifth HHG response. **b**, Illustration of twist-phase-matching during the propagation of a fundamental excitation wave and the generation of a fifth harmonic wave in a twisted hBN crystal. The P-B phase can be utilized to compensate for the phase mismatch. **c**, Twist-phase-matching condition of $t$ and $\theta$. **d**, Complex amplitude diagram illustrating the superposition of the fifth harmonic electric fields for twisted hBN flakes. **e**, Fifth HHG spectra of hBN flakes. Under 1340 nm pulsed laser excitation, the HHG signal of the assembled phase-matched hBN flakes with a twist angle of 30° is four times stronger than that of a single hBN flake, whereas the phase-mismatched hBN flakes with a twist angle of 0° exhibit a suppressed response.

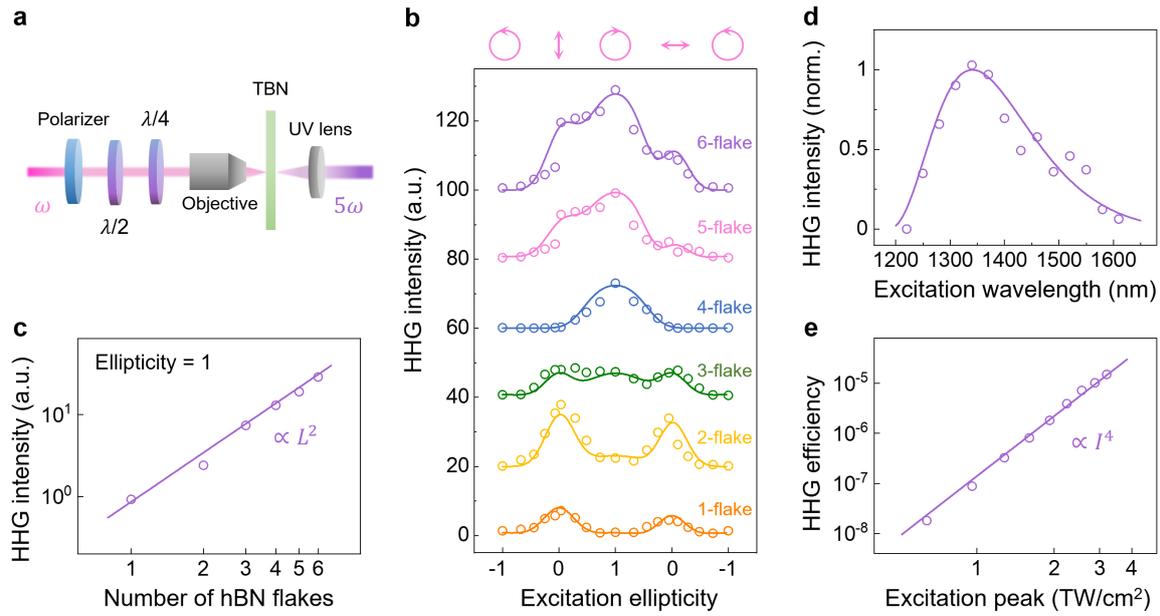

**Fig. 4 | HHG responses of multiple hBN flakes with a twist-phase-matching design. a**, Schematic of the experimental setup, showing the placement of optical components. $\lambda/2$ ($\lambda/4$), half-wave plate (quarter-wave plate). **b**, Theoretical (solid curves) and experimental (dots) results of the fifth HHG intensity from hBN flakes as a function of the excitation ellipticity and flake number. **c**, Flake number-dependent HHG intensity under ellipticity = 1. A quadratic dependence of the output HHG signals on the number of flakes is observed, indicating that the phase-matching condition is well satisfied. **d**, Excitation wavelength-dependent HHG intensity. The assembled six-flake hBN crystal has a phase-matching wavelength of 1340 nm and a bandwidth of ~200 nm. **e**, Pump-peak-intensity-dependent HHG conversion efficiency. The measured efficiency yield with a power law of 4 indicates a perturbative response under our experimental conditions. The maximum efficiency can reach 0.0015%.